\documentclass[a4paper,twocolumn,11pt,accepted=2022-06-29]{quantumarticle}
\pdfoutput=1
\usepackage[utf8]{inputenc}
\usepackage[english]{babel}
\usepackage[T1]{fontenc}
\usepackage{amsmath}
\usepackage{hyperref}

\usepackage[numbers]{natbib}

\usepackage{tikz}
\usepackage{lipsum}

\usepackage{dcolumn}    
\usepackage{bm} 
\usepackage{graphicx}
\usepackage{enumitem}
\usepackage{latexsym}
\usepackage{amsfonts}   
\usepackage{amssymb}
\usepackage{array}      
\usepackage{epsfig}
\usepackage{braket} 
\usepackage{soul}
\usepackage{hyperref}

\newcommand{\s}{\hat{\sigma}}

\newcommand{\ra}{\rangle}
\newcommand{\la}{\langle}

\newcommand{\Tr}{\textrm{Tr}}

\newcommand{\fig}{Fig.~}

\newcommand{\cf} {cf.~}

\DeclareRobustCommand\openzero{\leavevmode\hbox{0\kern-.55em0}}

\begin{document}

\title{Statistical time-domain characterization of non-periodic optical clocks}

\author{Dario Cilluffo}
\affiliation{Institute of Theoretical Physics \& IQST, Ulm University, Albert-Einstein-Allee 11 89081, Ulm, Germany}
\affiliation{Universit$\grave{a}$  degli Studi di Palermo, Dipartimento di Fisica e Chimica - Emilio Segr\`e, via Archirafi 36, I-90123 Palermo, Italy}
\orcid{0000-0001-6862-0511}

\begin{abstract}
Measuring time means counting the occurrence of periodic phenomena. Over the past centuries a major effort was put to make stable and precise oscillators to be used as clock regulators. Here we consider a different class of clocks based on stochastic clicking processes. We provide a rigorous statistical framework to study the performances of such devices and apply our results to a single coherently driven two-level atom under photodetection as an extreme example of non-periodic clock. Quantum Jump MonteCarlo simulations and photon counting waiting time distribution will provide independent checks on the main results.
\end{abstract}

We rely on physical systems that exhibit some kind of periodic behaviour for keeping accurate time: regularly repeated phenomena, which we will call \textit{events}, are recorded, the unit of time is thus defined as the time elapsed between two consecutive events and gets added to the count as soon as a new event is observed.
This is the principle behind the first clocks based on the observation of the motion of stars and planets.
Notwithstanding the devices used as clocks deviate from this basic idea due to unavoidable dissipation. 

\begin{figure}
\centering
\includegraphics[scale=0.40,angle=0]{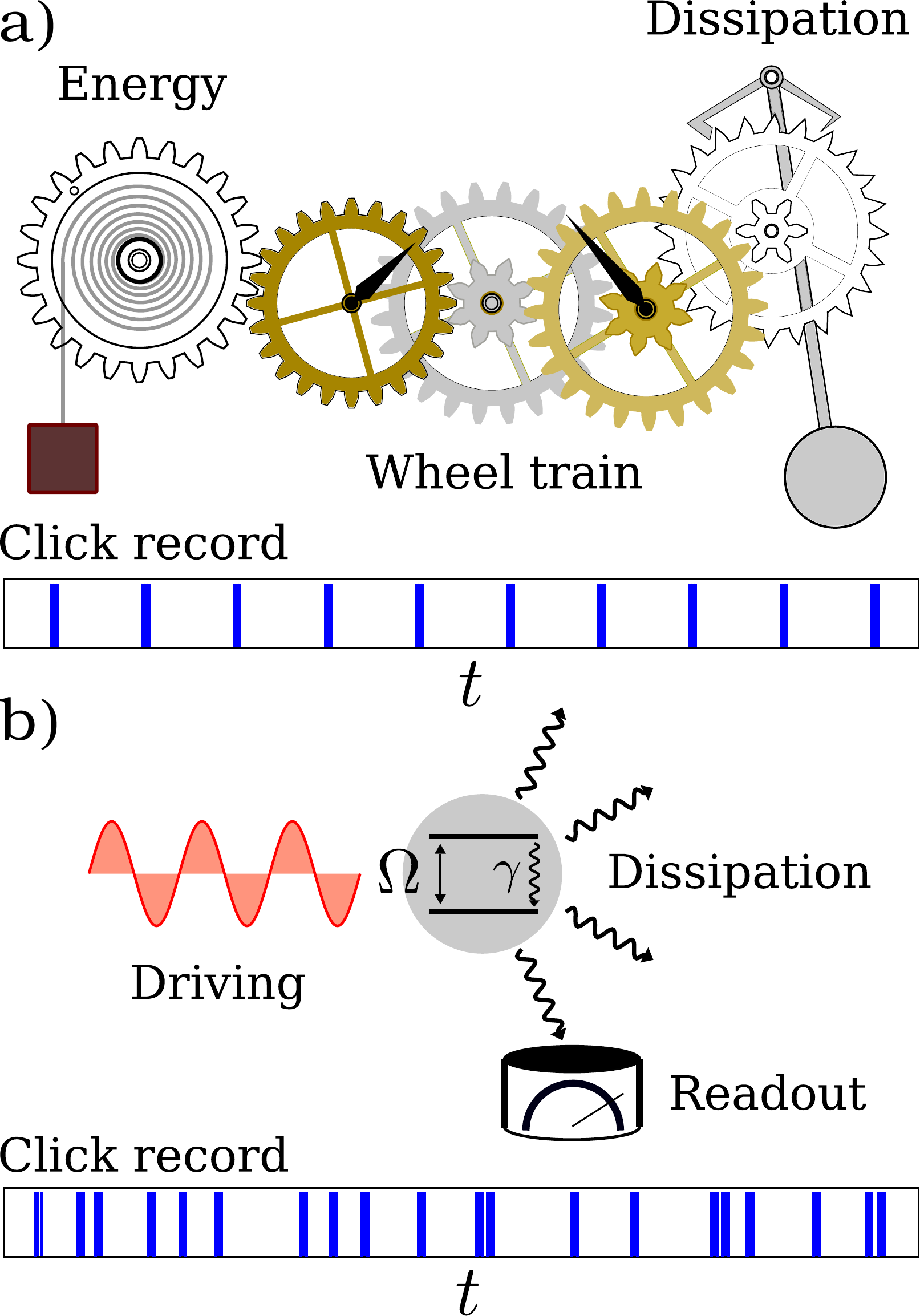}
\caption{(a) The most common mechanical clocks are composed of an oscillating device controlling the motion of a series of gears through which we read the time.
In the case of the pendulum here represented, the dead-beat escapement mechanism compensates the losses of energy due to friction and sustains the motion of the pendulum.
(b) The same principles apply to quantum optical systems used as clocks: the energy provided by a resonant driving field of frequency $\Omega$ is absorbed from the system (here a single two-level atom) and dissipated by emission with rate $\gamma$. A single photodetection event corresponds to a click.
}
\label{fig1}
\end{figure}
The best example are mechanical clocks: pendulums or hairsprings would stop after few oscillations due to friction without an escapement mechanism 
that provides the energy needed to sustain the motion. 
The energy is usually stored in a suspended weight or in a coiled spring during the winding operation, then it is transmitted to the swinging element through a wheel train (see \fig\ref{fig1}(a)).
Dissipation of energy implies unavoidable noise \cite{FordDiss} that introduces fluctuations in the operating frequency, making even a good clock device an inherently non-periodic system.
In the case of a dead-beat escapement clock like the one represented in \fig\ref{fig1}(a), fluctuations in the swinging amplitude of the pendulum affect the running rate of the wheel train, leading to error in timekeeping \cite{Mallock1911,kesteven1978mathematical}. A major issue in horology is quantifying such fluctuations over a clock working cycle and reducing them, and linear response theory provides good estimates in this and many other cases of practical interest \cite{hoyng2014dynamics,Ghosh_quartz}.
If we assume to deal with independent fluctuations, the central limit theorem allows to estimate straightforwardly the uncertainty over a measured time interval.
At the other extreme, we find timekeeping methods based on setups undergoing irreversible dynamics 
where fluctuations cannot directly be controlled
but, in special circumstances, turn out to be good clocks: that is the case of radiometric dating.
These are example of clocks in which statistics compensates for the lack of periodicity.
In recent works \cite{Milburn_clock,Erker2017PRX,QuantumWoods2021,PhysRevX.11.021029} such systems are studied from a thermodynamic viewpoint.
Here we perform an analytical study of the statistics of the events occurring in systems evolving through Markovian irreversible dynamics \cite{breuerTheory2007,  WisemanMilburnBook, Haroche,GardinerBook2004} by exploiting the formalism of quantum trajectories \cite{BrunLong, Brun}.
Through the study of this worst-case scenario, our main purpose is to offer a well-grounded framework within which the performances of non-periodic quantum clocks can be unambiguously defined and compared.
The possibility of studying analytically the stochastic fluctuations occurring in time measurement promises to be crucial in optimising also the performances of conventional clocks.

{\it Photon counting-based clocks.-}
Let us consider a measurement process over time yielding as outcome a signal consisting in a realization of a point process.
We will refer to the time-record of this signal as a \textit{trajectory}. Note that we are assuming to have access to a reference time $t$.
Let $N(t)$ be the number of events, or clicks, recorded at the reference time $t$ and $\mathcal{R}$ their rate. Thus the quantity
\begin{align}
\tau(t) = \frac{N(t)}{\mathcal{R}}\,,
\label{timedef}
\end{align}
represents the estimated elapsed time from a start time $t=0$. 
It is worth pointing out that if the considered clicking process serves as a good clock, one would expect that $\tau$ must be close to $t$, by definition. 

The evolution of a quantum system under continuous monitoring is characterized by quantum jumps:
the measurement process makes the wavefunction describing the system collapses at some times. Thus some specific observables will feature a point-process behaviour in time, known as \textit{quantum trajectory} \cite{Brun}. We use collapse events to mark the time according to \eqref{timedef}.
The goal is to find a sufficiently simple expression of the error $\delta \tau$ and a minimisation strategy.
In the particular case of optical damped systems subject to photodetection it proves to be a very easy task.

\begin{figure}	
\includegraphics[scale=0.7,angle=0]{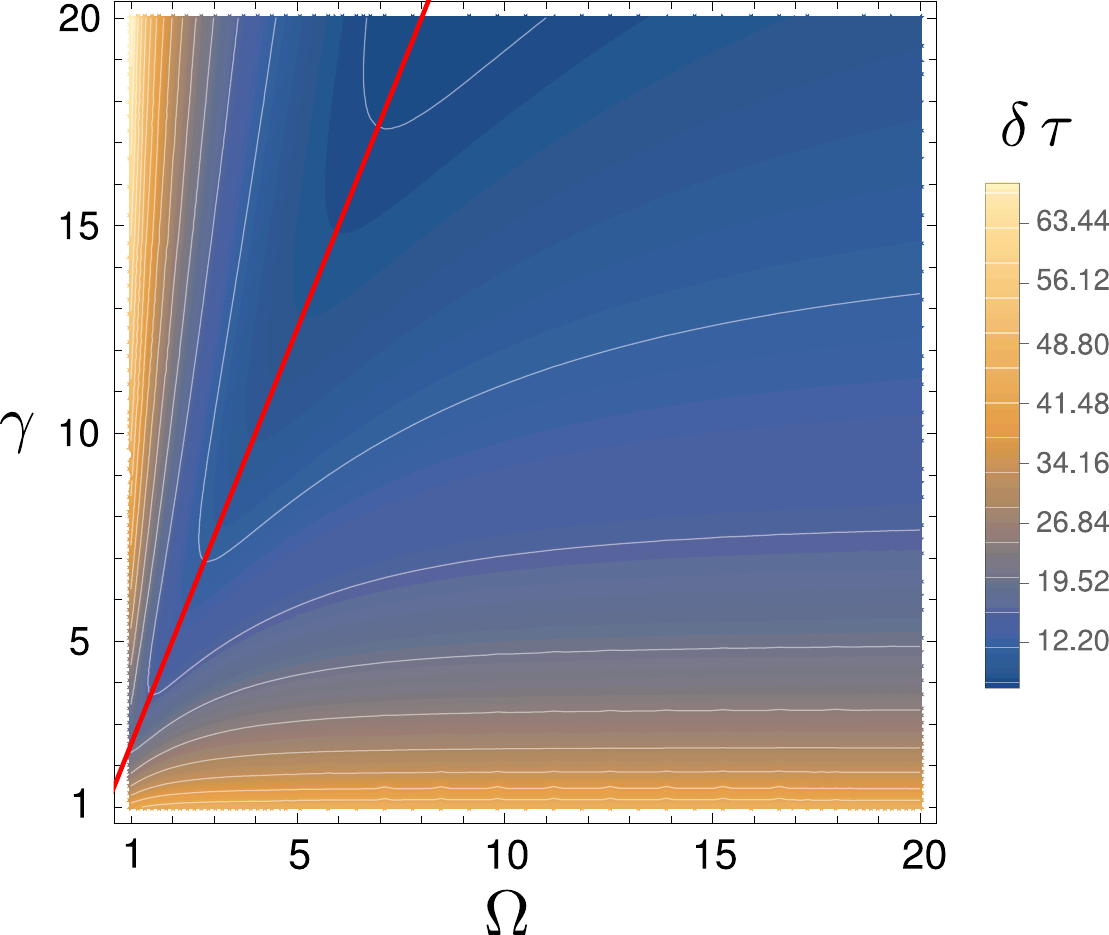}
\caption{$\delta \tau$ as a function of $\gamma$ and $\Omega$ for a two-level atom in resonance with a coherent driving. We fixed $t=1000$. The red minimum line corresponds to $\gamma=\tfrac{5}{2} \Omega$.}
\label{figerr}
\end{figure}

Assume, for simplicity, that the system and the environment can exchange excitations (photons) through a single channel and we study their evolution over the time interval $\delta t$. The coupling strength between system and environment is assumed to be very small compared with the typical energy scales of the system free Hamiltonian $\hat{H}$ (weak coupling). 
The loss of a quantum by the system corresponds to a sudden change
in its state $\rho$ due to the action of a specific system's Hilbert space operator $\hat{c}$ (\textit{jump operator}). Thus in this case we have
\begin{align}
\delta \rho = \frac{\hat{c} \rho \hat{c}^\dag}{\Tr\{ \hat{c}^\dag  \hat{c} \rho \}} -\rho  = \mathcal{J}[\rho]  \, .
\label{ev1}
\end{align}
We define the variable $\delta N_c(t)$ as the number of quanta the system can lose over $\delta t$. In the case of evolution described by \eqref{ev1} we have $\delta N_c(t) = 1$.
If there are no jumps ($\delta N_c(t)=0$) the system evolves under the effective non-Hermitian Hamiltonian $\hat{H}_{\rm eff}=\hat{H}-\tfrac{i}{2} \hat{c}^\dag\hat{c}$ \cite{Brun,Haroche,PlenioJumpRMP} 
\begin{align}
\delta \rho &= \frac{e^{-i \hat{H}_{\rm eff} \delta t } \, \rho \, e^{i \hat{H}_{\rm eff}^\dag \delta t }}{\Tr\{ e^{i \hat{H}_{\rm eff}^\dag \delta t } \,  e^{-i \hat{H}_{\rm eff} \delta t } \, \rho \}} \notag
\\&  \simeq  -i [\hat{H}_{\rm eff} \rho - \rho \hat{H}_{\rm eff}^\dag] \delta t + \rho \Tr\{ \hat{c}^\dag  \hat{c} \rho \} \delta t = \delta t \, \mathcal{H}[\rho]
\,,
\label{ev2}
\end{align}
where the last expression holds for $\delta t$ very short compared to the characteristic evolution time scale of the environment \cite{Haroche}.
By combining \eqref{ev1} and \eqref{ev2} through the process $\delta N_c(t)$
the evolution of the system conditioned on previous measurement results obeys the stochastic master equation
\begin{align}
d \rho_c (t) = \left( \delta N_c(t) \mathcal{J} - \delta t \, \mathcal{H} \right) [\rho_c (t)]\,,
\label{SME}
\end{align}
where the subscript $c$ is for ``conditioned". Note that the variable
$d N_c(t)$, which is a stochastic point process taking only the values $0$ and $1$ with average $\langle d N_c\rangle = \Tr\{ \hat{c}^\dag  \hat{c} \rho \} \delta t$ and variance $\delta N_c^2 = \delta N_c$ \cite{WisemanMilburnBook},
describes the photon counting process and provides the signal (\textit{emission trajectory}) we can use to mark the time.
For each trajectory we can define an estimated time $\tau_c (t)$ according to \eqref{timedef}:
\begin{align}
\tau_c (t) \propto \int_0^t d t' d N_c(t)\,.
\label{ave}
\end{align}
We are interested in the distribution of such times, in particular in their typical values.
This requires switching to the unconditioned dynamics as it will be clear below. 
\begin{figure*}
\includegraphics[scale=0.45,angle=0]{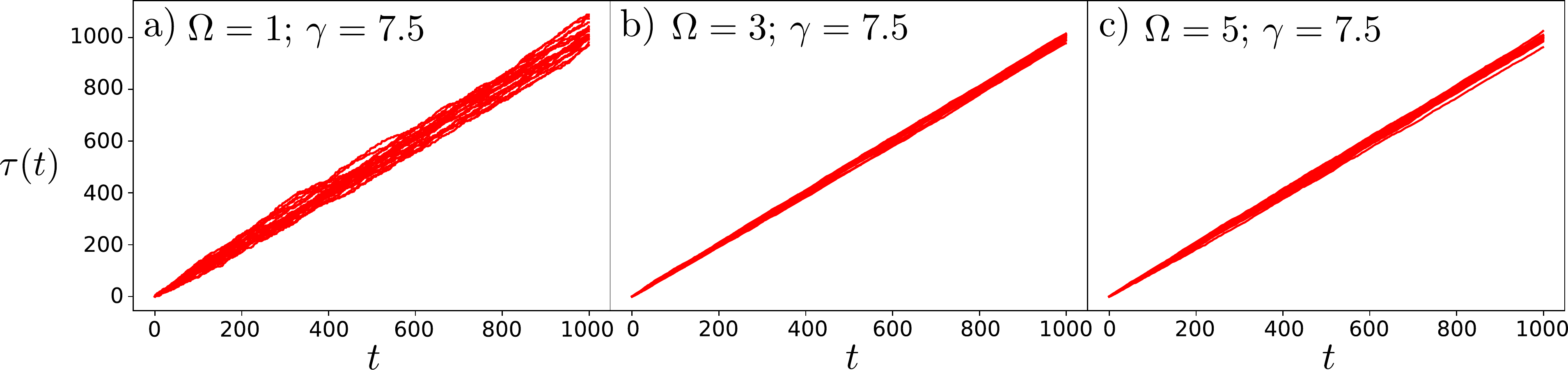}
\caption{Simulated measurement of time through a coherently-driven decaying two-level atom 
in three different regimes of parameters ($\Omega$ and $\gamma$).
In each plot there are $20$ lines each corresponding to a QJMC run. In particular parameters in (b) lie on the line $\gamma=5/2 \, \Omega$. 
}
\label{fig4}
\end{figure*}
\\

{\it Asymptotic rate in the Markovian regime.-}
The unconditioned state of the system at time $t$ is recovered by averaging over all the possible results of the measurement, i.e.~over the whole ensemble $\mathcal{N}_t$ of quantum trajectories at time $t$ \cite{manzano2014symmetry, belokurov2002conditional}
\begin{align}
\rho(t) = \la \rho_c(t) \ra_{\mathcal{N}_t} \, ,
\end{align}
and under the Markov condition the stochastic master equation \eqref{SME} is turned into the celebrated GKSL master equation (ME) \cite{Haroche,gorini1976completely,lindblad1976generators}
\begin{align}
\dot{\rho} = \mathcal{L} \rho = - i [\hat{H} , \rho] + \mathcal{D}[\hat{c}]\rho 
\label{GKSL}
\end{align}
where we defined the dissipator $\mathcal{D}[\hat{A}]\rho = \hat{A}\rho \hat{A}^\dagger - \tfrac{1}{2} \{\hat{A}^\dag \hat{A},\rho\}$.
The statistical ensemble $\mathcal{N}_t$ is characterized by the moment generating function
\begin{align}
Z_t(s) = \sum_{N=0}^\infty P_t(N) e^{- s N}\, ,
\end{align}
where $P_t(N)$ is the probability to observe $N$ counts after the time $t$. In the limit of large times, i.e.~for $t$ very large with respect to the characteristic timescales of the system, it can be approximated with the \textit{large deviation} form \cite{ellis1996overview,Touchette,Vulpiani}
\begin{align}
Z_t(s) \simeq e^{t \theta(s)}\, ,
\end{align}

where the the time-scaled cumulant generating function $\theta(s)=\tfrac{1}{t} \log (Z_t(s))$ is easily recovered by exact diagonalization as the maximum real eigenvalue of the \textit{biased} superoperator $\mathcal{L}_s$ \cite{IgorPRL2010,Touchette}
\begin{align}
\mathcal{L}_s \rho = - i [\hat{H} , \rho] + e^{-s} \mathcal{D}[\hat{c}]\rho \, .
\end{align}
Once the moment generating function is known, we can access all the features of the statistics of trajectory by direct derivation \cite{JordanBook1965}.
In particular
\begin{align}
\mathcal{R} = \frac{\la N \ra_t}{t} = -\partial_s\theta(s) \bigg|_{s=0}\,\!\!\!\!.
\label{apprld}
\end{align}
It is worth recalling here that most of the commonly used clock devices are machines equipped with a feedback system controlling their operating speed (e.g. the pendulum in mechanical clocks). 
Achieving a balance through the action of the feedback control generally takes a certain number of cycles, during which the machine is not reliable.
Thus the use of asymptotic estimates for the moments of $P_t$ is justified since we usually require a clock to work well in stationary conditions and we are not interested in transient steps after turning it on.

From here we will adopt the Lagrange (primed) notation for partial derivatives with respect to $s$ and we will omit the argument $s$ when it is zero, thus $\mathcal{R} = -\theta'$.
Analogously it turns out that $\theta'' = \tfrac{1}{t}(\la N^2 \ra_t - \la N \ra_t^2)$ so that, in the limit of long times $t$, we have
\begin{align}
N(t) = -\theta' \, t \pm \sqrt{\theta'' \, t}\, ,
\end{align}
that, dividing by the rate $\mathcal{R}$, returns the estimated time 
\begin{align}
\tau (t) = t \pm \frac{\sqrt{\theta'' \, t}}{\theta'}\,.
\label{key}
\end{align}

This simple equality is the key result of this work.
We could have achieved a similar result by studying the time intervals to be added to the count as i.i.d.~stochastic variables through a direct application of the central limit theorem. 
This however requires a deep analysis of the waiting time distribution of the photon counting process, which, except in particular cases, can be nontrivial to work out in explicit analytic form \cite{Vacchini_2014}.
Note that the absolute error increases with the square root of the time $t$, so the effects of ”faults” worsen the reliability of the estimated time, but the relative error decreases
$(\propto t^{-1/2})$.

In the remaining we will study the performances of a two-level atom driven by a resonant classical source at zero temperature under direct photodetection used as a clock.
This setup has already been studied in details from the point of view of quantum trajectories \cite{IgorPRL2010} and has the advantage of not requiring advanced measurement techniques in the optical regime.
Despite, as it will result from the discussion, the quality of the clock signal provided by this simplified system is not particularly good,
it allows us to test the formalism in a model whose behaviour is well known.
\\

{\it Coherently driven two-level atom.-}
We consider a two-level atom with energy separation $\Omega$ driven by a classical resonant field and decaying into a vacuum environment with rate $\gamma$ (\cf\fig\ref{fig1}-b).
In these conditions the master equation \eqref{GKSL} describing the reduced unconditioned dynamics of the system reads
\begin{align}
\dot{\rho} = \mathcal{L} \rho = - i [\Omega (\s_+ + \s_-) , \rho] + \mathcal{D}[\sqrt{\gamma} \s_-]\rho \, .
\end{align}
Exact diagonalization of $\mathcal{L}$ returns the cumulant generating function:
\begin{align}
&\theta(s) = \frac{1}{2} \left(-\gamma  + \frac{e^{-4 s}}{\sqrt[3]{3^2}} \,\mathcal{C} + \frac{e^{4 s}}{\sqrt[3]{3}} \frac{\mathcal{A}}{\mathcal{C}} \right)\,,
\end{align}
with
\begin{align}
&\mathcal{A} = \gamma ^2-16 \, \Omega ^2 \,,\\ 
&\mathcal{B} = 72 \gamma  \,\Omega ^2 e^{11 s} \,,\\
&\mathcal{C}=\sqrt[3]{\sqrt{ \mathcal{B}^2 -3 \, e^{24 s} \mathcal{A}^3}+\mathcal{B}} \,.
\end{align}
The analytic expression of the cumulants are not reported here for convenience.
In \fig\ref{figerr} we report the error $\delta \tau  = \tfrac{\sqrt{\theta'' \, t}}{\theta'}$ as a function of the driving frequency and the decay rate, for $t=1000$. An higher dissipation corresponds to a smaller error, as expected for dissipative systems used as clocks \cite{Milburn_clock}. 
At small values of $\Omega$ the system is overdamped and re-excitation after decay becomes very rare, resulting in increasing fluctuations of the time intervals between two consecutive emission events.
Note that $\delta \tau$ reaches its minimum on the line $\gamma=\tfrac{5}{2}\Omega$.

\begin{figure}	
\includegraphics[scale=0.7,angle=0]{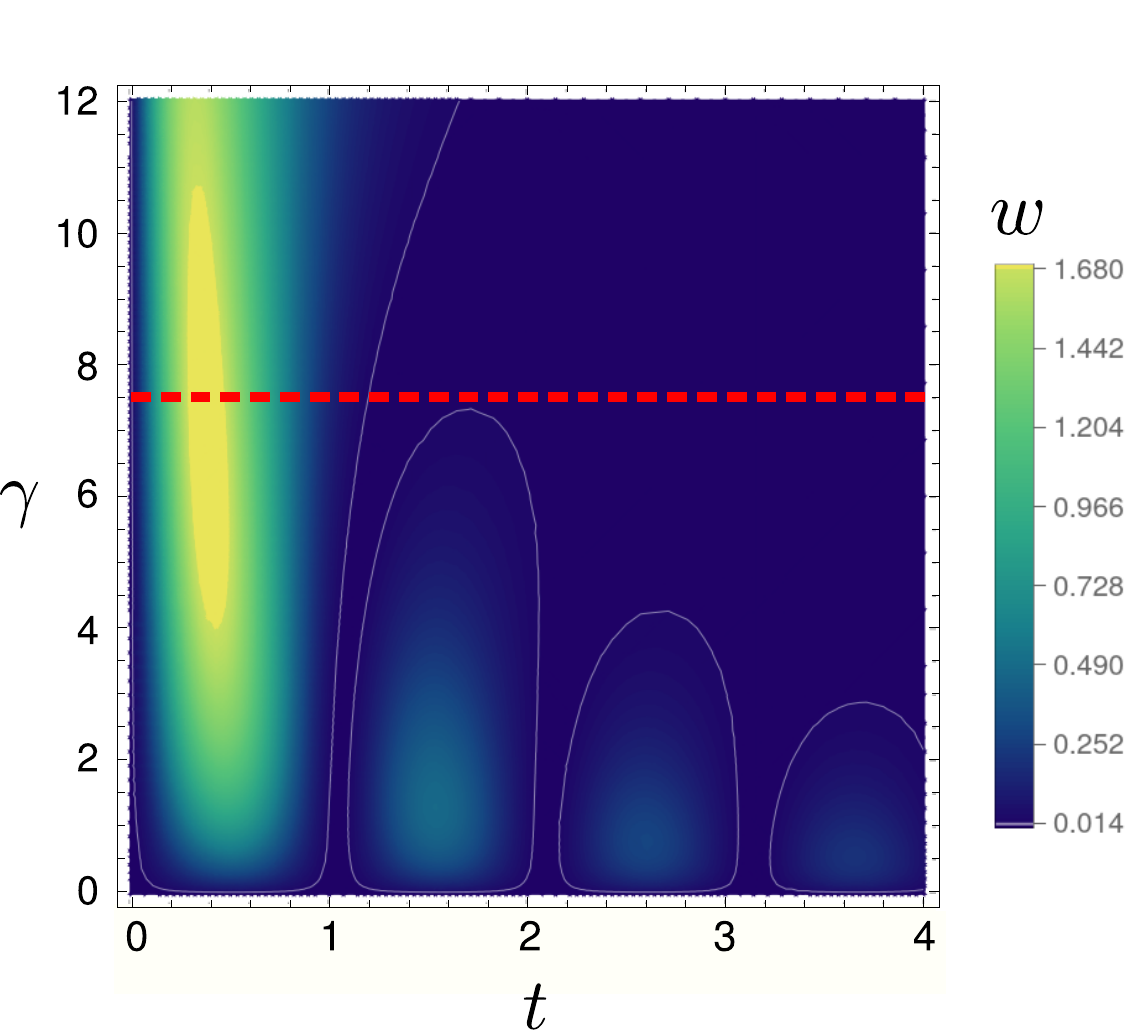}
\caption{Waiting time distribution (\cf \eqref{wtdeq}) as a function of $\gamma$ with $\Omega=3$. The red dashed line corresponds to $\gamma=\tfrac{5}{2} \Omega$. The peaks are marked through the white contours ($w \simeq 0.014$).}
\label{figwtd}
\end{figure}

This behaviour is reflected by simulated trajectories in \fig\ref{fig4}. Using Quantum Jump MonteCarlo (QJMC) \cite{carmichaelbook2009,PlenioJumpRMP, Haroche} we generated $20$ trajectories for each set of parameters.
Each red line represents a different realization of $\tau(t)$ (\cf\eqref{timedef}) calculated from a single quantum trajectory.
It can be noted how the lines spread along time and the error increases, as expected by \eqref{key}. The parameters in \fig\ref{fig4}-b lie on the minimum-error line and this corresponds to the lowest spreading, i.e.~the clock turns out to be more accurate.

Since in this setup for any $\rho$ it holds $\gamma \, \s_- \rho \s_+ = |g\ra \la g| \, \Tr\{ \s_- \rho \s_+ \}$, we can straightforwardly work out the waiting time distribution of the emission process \cite{CarmichaelPRA89,Vacchini_2014} as a function of $\Omega$ and $\gamma$
\begin{align}
w(\Omega, \gamma, t) = 
\gamma \, \Omega^2 e^{-t \gamma /2} 
\left[\frac{ \sinh\left( \frac{\sqrt{\mathcal{A} }}{4} t  \right) }{\left(\frac{\sqrt{\mathcal{A}}}{4}\right)}\right]^2\, .
\label{wtdeq}
\end{align}
The analysis of the waiting time distribution provides a clear interpretation of the line of minima and of the behaviour of $\delta\tau$.
In \fig\ref{figwtd} we show the waiting time distribution for $\Omega = 3$. The red dashed line corresponds to $\tilde{\gamma}=\tfrac{5}{2} \Omega = 7.5$. For $\gamma<\tilde{\gamma}$ the waiting time distribution features a main peak and several satellite peaks, making the time interval between two clicks unpredictable. At $\gamma=\tilde{\gamma}$ only the main peak survives and the uncertainty over click time is strongly reduced, but as $\gamma$ increases the peak width grows, resulting in a less accurate measure of the time, in agreement with the large deviation estimate (see \fig\ref{figerr}).
It is worth noting that even in the best case scenario the accuracy of this particular clock is not very high, as it can be directly estimated from the WTD: the width of the central peak is $\sigma \sim 1s$ and the average interval $\Delta t \sim 1/2 s$, thus the number of intervals we need to cover the total elapsed time ($1000s$) amounts to $n\sim 2 \cdot 10^3$.
If we roughly approximate the peak with a Gaussian distribution we expect $\sqrt{n}\cdot \sigma=40 s$ as standard deviation of the distribution of the total elapsed time, that is of the same order of $3 \, \delta\tau = 30s$ given by \eqref{key} in the same regime.
\\

{\it Conclusions.-}
As dissipative physical systems, clocks are constrained by the known laws of thermodynamics that put fundamental limitations on their performances.
An investigation into these aspects, in particular in the case of quantum systems, goes through the study of the entropy balance of the measurement processes involved in timekeeping (measurement driven clocks \cite{Milburn2021}).
Our point of view is different but complementary: we regard the clock signals as microstates within a statistical ensemble. We have shown how thermodynamics of quantum trajectories provides a new way to investigate the performances of non-periodic clocks implemented through simple optical quantum systems.

Through the extreme example of a coherently driven two-level atom we have shown that our results perfectly match with the information provided by the waiting time 
distribution and quantum MonteCarlo simulations.
Although, as pointed out at the end of the last section, the simplified model here studied does not provide an alternative to currently used clocks,
our results suggest that the large deviation approach looks promising for making good predictions about the performances of clocks and may be extended to other unravellings \cite{Hickey_2012,CilluffoJSTAT2019} and, more in general, a wide variety of dissipative dynamics that could realistically be exploited to produce a clock signal on many experimental platforms.
In particular photon counting measurement with high detection efficiency in the optical wavelengths is a non-demanding task with current technology and lacks in detection efficiency can be easily modelled within the framework here presented through additional decay channels.
Moreover, the knowledge of the function $\theta(s)$ enables us to access the higher-order moments of the asymptotic distribution $P_t(N)$ and therefore also to draw information that are not covered by the central limit theorem \cite{Touchette}.
Note that in our calculation we limited to the time average \eqref{ave} as estimator of the elapsed time, but in principle other estimator could yield better clock signals. In future research it will be worth investigating the meeting points of our large deviation approach and the proper tools of quantum metrology.
In principle this formalism can be applied also to more conventional and realistic clocks in order to describe small inaccuracies in time measurement due to the unavoidable deviations from perfect periodicity, especially in cases in which an analytical study of the 
waiting time distribution of the counting process is difficult.
\\

\textit{Acknowledgements.-}
The author would like to thank F. Ciccarello, S. Lorenzo, G. M. Palma, A. Carollo, D. A. Chisholm and C. Pellitteri for fruitful discussions.
The author acknowledges support from MIUR through project PRIN Project 2017SRN-BRK QUSHIP.

\bibliographystyle{quantum}
\bibliography{biblio}

\begin{thebibliography}{10}

\bibitem{FordDiss}
G.~W. {Ford}.
\newblock ``{The fluctuation-dissipation theorem}''.
\newblock \href{https://dx.doi.org/10.1080/00107514.2017.1298289}{Contemporary
  Physics {\bf 58}, 244--252}~(2017).

\bibitem{Mallock1911}
Henry Reginald~Arnulph Mallock.
\newblock ``Pendulum clocks and their errors''.
\newblock \href{https://dx.doi.org/10.1098/rspa.1911.0064}{Proceeding of the
  Royal Society A{\bf 85}}~(1911).

\bibitem{kesteven1978mathematical}
M~Kesteven.
\newblock ``On the mathematical theory of clock escapements''.
\newblock \href{https://dx.doi.org/}{American Journal of Physics {\bf 46},
  125--129}~(1978).

\bibitem{hoyng2014dynamics}
Peter Hoyng.
\newblock ``Dynamics and performance of clock pendulums''.
\newblock \href{https://dx.doi.org/}{American Journal of Physics {\bf 82},
  1053--1061}~(2014).

\bibitem{Ghosh_quartz}
S.~Ghosh, F.~Sthal, J.~Imbaud, M.~Devel, R.~Bourquin, C.~Vuillemin, A.~Bakir,
  N.~Cholley, P.~Abbe, D.~Vernier, and G.~Cibiel.
\newblock ``Theoretical and experimental investigations of 1/f noise in quartz
  crystal resonators''.
\newblock \href{https://dx.doi.org/10.1109/EFTF-IFC.2013.6702262}{2013 Joint
  European Frequency and Time Forum International Frequency Control Symposium
  (EFTF/IFC)Pages 737--740}~(2013).

\bibitem{Milburn_clock}
G.~J. Milburn.
\newblock ``The thermodynamics of clocks''.
\newblock \href{https://dx.doi.org/10.1080/00107514.2020.1837471}{Contemporary
  Physics {\bf 61}, 69--95}~(2020).

\bibitem{Erker2017PRX}
Paul Erker, Mark~T. Mitchison, Ralph Silva, Mischa~P. Woods, Nicolas Brunner,
  and Marcus Huber.
\newblock ``Autonomous quantum clocks: Does thermodynamics limit our ability to
  measure time?''.
\newblock \href{https://dx.doi.org/10.1103/PhysRevX.7.031022}{Phys. Rev. X {\bf
  7}, 031022}~(2017).

\bibitem{QuantumWoods2021}
Mischa~P. Woods.
\newblock ``Autonomous ticking clocks from axiomatic principles''.
\newblock \href{https://dx.doi.org/10.22331/q-2021-01-17-381}{Quantum {\bf 5},
  381}~(2021).

\bibitem{PhysRevX.11.021029}
A.~N. Pearson, Y.~Guryanova, P.~Erker, E.~A. Laird, G.~A.~D. Briggs, M.~Huber,
  and N.~Ares.
\newblock ``Measuring the thermodynamic cost of timekeeping''.
\newblock \href{https://dx.doi.org/10.1103/PhysRevX.11.021029}{Phys. Rev. X
  {\bf 11}, 021029}~(2021).

\bibitem{breuerTheory2007}
Heinz-Peter Breuer and Francesco Petruccione.
\newblock ``The theory of open quantum systems''.
\newblock
  \href{https://dx.doi.org/10.1093/acprof:oso/9780199213900.001.0001}{Oxford
  University Press}. ~(2007).

\bibitem{WisemanMilburnBook}
Howard~M. Wiseman and Gerard~J. Milburn.
\newblock ``{Quantum measurement and control}''.
\newblock \href{https://dx.doi.org/10.1017/CBO9780511813948}{Volume
  9780521804424, pages 1--460}.
\newblock Cambridge university press. ~(2009).

\bibitem{Haroche}
Serge Haroche and Jean~Michel Raimond.
\newblock ``{Exploring the Quantum: Atoms, Cavities, and Photons}''.
\newblock
  \href{https://dx.doi.org/10.1093/acprof:oso/9780198509141.001.0001}{Oxford
  Univ. Press}. Oxford~(2006).

\bibitem{GardinerBook2004}
Crispin Gardiner, Peter Zoller, and Peter Zoller.
\newblock ``Quantum noise: a handbook of markovian and non-markovian quantum
  stochastic methods with applications to quantum optics''.
\newblock \href{https://dx.doi.org/10.48550/ARXIV.QUANT-PH/9702030}{Springer
  Science \& Business Media}. ~(2004).

\bibitem{BrunLong}
Todd~A. Brun.
\newblock ``Continuous measurements, quantum trajectories, and decoherent
  histories''.
\newblock \href{https://dx.doi.org/10.1103/physreva.61.042107}{Physical Review
  A{\bf 61}}~(2000).

\bibitem{Brun}
Todd~A. Brun.
\newblock ``A simple model of quantum trajectories''.
\newblock \href{https://dx.doi.org/10.1119/1.1475328}{American Journal of
  Physics {\bf 70}, 719–737}~(2002).

\bibitem{PlenioJumpRMP}
M.~B. Plenio and P.~L. Knight.
\newblock ``The quantum-jump approach to dissipative dynamics in quantum
  optics''.
\newblock \href{https://dx.doi.org/10.1103/RevModPhys.70.101}{Rev. Mod. Phys.
  {\bf 70}, 101--144}~(1998).

\bibitem{manzano2014symmetry}
Daniel Manzano and Pablo~I Hurtado.
\newblock ``Symmetry and the thermodynamics of currents in open quantum
  systems''.
\newblock
  \href{https://dx.doi.org/https://doi.org/10.1103/PhysRevB.90.125138}{Phys.
  Rev. B {\bf 90}, 125138}~(2014).

\bibitem{belokurov2002conditional}
VV~Belokurov, OA~Khrustalev, VA~Sadovnichy, and OD~Timofeevskaya.
\newblock ``Conditional density matrix: Systems and subsystems in quantum
  mechanics''~(2002).
\newblock
  url:~\href{https://arxiv.org/abs/quant-ph/0210149}{arxiv.org/abs/quant-ph/0210149}.

\bibitem{gorini1976completely}
Vittorio Gorini, Andrzej Kossakowski, and Ennackal Chandy~George Sudarshan.
\newblock ``Completely positive dynamical semigroups of n-level systems''.
\newblock \href{https://dx.doi.org/https://doi.org/10.1063/1.522979}{Journal of
  Mathematical Physics {\bf 17}, 821--825}~(1976).

\bibitem{lindblad1976generators}
Goran Lindblad.
\newblock ``On the generators of quantum dynamical semigroups''.
\newblock
  \href{https://dx.doi.org/https://doi.org/10.1007/BF01608499}{Communications
  in Mathematical Physics {\bf 48}, 119--130}~(1976).

\bibitem{ellis1996overview}
RS~Ellis.
\newblock ``An overview of the theory of large deviations and applications to
  statistical mechanics.''.
\newblock
  \href{https://dx.doi.org/https://doi.org/10.1080/03461238.1995.10413952}{Insurance
  Mathematics and Economics {\bf 3}, 232--233}~(1996).

\bibitem{Touchette}
Hugo Touchette.
\newblock ``The large deviation approach to statistical mechanics''.
\newblock \href{https://dx.doi.org/10.1016/j.physrep.2009.05.002}{Physics
  Reports {\bf 478}, 1--69}~(2009).

\bibitem{Vulpiani}
Angelo Vulpiani, Fabio Cecconi, Massimo Cencini, Andrea Puglisi, and Davide
  Vergni.
\newblock ``Large deviations in physics''.
\newblock
  \href{https://dx.doi.org/https://doi.org/10.1007/978-3-642-54251-0}{The
  Legacy of the Law of Large Numbers (Berlin: Springer)}~(2014).

\bibitem{IgorPRL2010}
Juan~P Garrahan and Igor Lesanovsky.
\newblock ``Thermodynamics of quantum jump trajectories''.
\newblock \href{https://dx.doi.org/10.1103/physrevlett.104.160601}{Phys. Rev.
  Lett. {\bf 104}, 160601}~(2010).

\bibitem{JordanBook1965}
Charles Jordan and K{\'a}roly Jord{\'a}n.
\newblock ``Calculus of finite differences''.
\newblock \href{https://dx.doi.org/}{Volume~33}.
\newblock American Mathematical Soc. ~(1965).

\bibitem{Vacchini_2014}
Bassano Vacchini.
\newblock ``General structure of quantum collisional models''.
\newblock \href{https://dx.doi.org/10.1142/s0219749914610115}{International
  Journal of Quantum Information {\bf 12}, 1461011}~(2014).

\bibitem{carmichaelbook2009}
Howard Carmichael.
\newblock ``An open systems approach to quantum optics: lectures presented at
  the universit{\'e} libre de bruxelles, october 28 to november 4, 1991''.
\newblock \href{https://dx.doi.org/}{Volume~18}.
\newblock Springer Science \& Business Media. ~(2009).

\bibitem{CarmichaelPRA89}
H.~J. Carmichael, Surendra Singh, Reeta Vyas, and P.~R. Rice.
\newblock ``{Photoelectron waiting times and atomic state reduction in
  resonance fluorescence}''.
\newblock \href{https://dx.doi.org/10.1103/PhysRevA.39.1200}{Physical Review A
  {\bf 39}, 1200--1218}~(1989).

\bibitem{Milburn2021}
A.~A. Gangat and G.~J. Milburn.
\newblock ``Quantum clocks driven by measurement''~(2021).
\newblock  \href{http://arxiv.org/abs/2109.05390}{arXiv:2109.05390}.

\bibitem{Hickey_2012}
James~M. Hickey, Sam Genway, Igor Lesanovsky, and Juan~P. Garrahan.
\newblock ``Thermodynamics of quadrature trajectories in open quantum
  systems''.
\newblock \href{https://dx.doi.org/10.1103/physreva.86.063824}{Physical Review
  A{\bf 86}}~(2012).

\bibitem{CilluffoJSTAT2019}
Dario Cilluffo, Salvatore Lorenzo, G~Massimo Palma, and Francesco Ciccarello.
\newblock ``Quantum jump statistics with a shifted jump operator in a chiral
  waveguide''.
\newblock \href{https://dx.doi.org/10.1088/1742-5468/ab371c}{Journal of
  Statistical Mechanics: Theory and Experiment {\bf 2019}, 104004}~(2019).

\end{thebibliography}

\end{document}